\documentstyle[psfig,rotate,12pt]{article}
\begin{document} 

\title{The energy dependence of the in-medium $\eta N$ cross section 
evaluated from $\eta$-photoproduction
\thanks{Supported by Forschungszentrum J\"ulich, GSI, BMBF and DFG}}
\author{
M. Effenberger and A. Sibirtsev\\
Institut f\"ur Theoretische Physik, Universit\"at Giessen \\
D-35392 Giessen, Germany}

\date{ }

\maketitle

\begin{abstract}
Within the Glauber formalism and a BUU transport model
we analyze the $\eta$-photoproduction data 
from nuclei and evaluate the in-medium $\eta N$ cross section.
Our results indicate that the $\eta N$ cross section 
is almost independent of the $\eta$ energy up to 200 MeV.
\end{abstract}

\section{Introduction}
For a long time $\eta$-meson production in nuclei
has been of interest as a source of
information about the $\eta$-nucleus final state 
interaction. 
The present knowledge about the $\eta N$ interaction even
in the vacuum comes from either simple analysis
of the inverse $\pi N \to \eta N$ reaction or as a
free parameter fitted to experimental data by theoretical 
calculations~\cite{Liu,Wilkin,Chiavassa,Sibirtsev1}.

Note that the value of the $\eta N$ scattering length is
still an open problem and there is not actual agreement
between a bulk of theoretical investigations.

The analysis of $\eta$-production from $pA$ collisions
indicated strong sensitivity of the calculations to the
prescription of the $\eta$-meson final state 
interaction~\cite{Chiavassa,Golubeva,Sibirtsev2}.
It was found that  the $\eta$-energy 
spectrum~\cite{Chiavassa2}  is
mostly influenced by the variation of the $\eta N$ cross 
section~\cite{Golubeva,Sibirtsev2}. However the experimental
data~\cite{Chiavassa2} had large uncertainties and there
was no continuation of the systematical studies. 

Recent measurements on $\eta$-meson photoproduction
in nuclei~\cite{Krusche} are more detailed and accurate. 
Among the theoretical
investigations~\cite{Lee,Carrasco,Effenberger} only the
calculations within the Distorted Wave Impulse
Approximation (DWIA) from Lee et al.\cite{Lee} are able to
reproduce the experimental data by incorporating
the $\eta$-nucleus potential proposed by
Bennhold and Tanabe~\cite{Ben}. 

The present paper is organized as follows. In section~\ref{glauber}
we use the Glauber formalism to extract the in-medium $\eta N$ cross
section from the experimental data. In section~\ref{buu} these
results are compared to Boltzmann-Uehling-Uhlenbeck transport 
model calculations. The sensitivity
of the theoretical results to the prescription of $\eta N$ scattering is
investigated.

\section{Analysis within the Glauber Model}
\label{glauber}
In an incoherent approximation the cross section of 
$\eta$-meson photoproduction off nuclei is given by
\begin{equation}
\label{app1}
{\sigma}_{\gamma \eta}^A = 
{\sigma}_{\gamma \eta }^p \times
\left[ Z  + \zeta (A-Z) \right] 
\end{equation}
with $A$, $Z$ being the mass and charge of the target,respectively, 
while the factor $\zeta=2/3$~\cite{Krusche2} 
stands for the ratio of the elementary 
$\eta$-photoproduction cross sections  from
$\gamma n$ and $\gamma p$ reactions.

In nuclei the cross section differs from the  
approximation~(\ref{app1}) due to nuclear effects. 
({\it  i})~The Fermi motion
of nucleons as well as ({\it  ii})~Pauli blocking are important
at energies below and close to the reaction threshold in 
free space~\cite{Cassing1,Salcedo}. We should also take into account 
{\it (iii)}~the modification of the $N^*$-resonance
by the nuclear medium~\cite{Carrasco,Effenberger}. 

However the most important
effect is {\it (iv)}~the strong final state interaction
of $\eta $-mesons in nuclear matter.
The deviation of the $A$-dependence of the 
${\sigma}_{\gamma A \rightarrow \eta X}$ from $A^1$ mostly
reflects the strength of the final state interaction.

Here we present  an analysis of the experimental
data on $\gamma A \rightarrow \eta X$ reactions in order to
extract the in-medium cross section ${\sigma}_{\eta N}$. 
Our approach is based on the Glauber model~\cite{Glauber} 
and first was developed by Margolis~\cite{Margolis1}
for evaluation of the $\rho N$ cross section from both
incoherent and coherent $\rho$-meson photoproduction off nuclei.
The most detailed description and application of the Glauber model
to photoproduction reactions may be found from review of Bauer, Spital 
and Yennie~\cite{Bauer}.
A similar formalism is adopted for studying  color 
transparency~\cite{Bertch,Kopeliovich1}, where the in-medium
cross sections is treated as a function of a transverse 
separation of the hadronic wave function.

In the Glauber model the cross section of the incoherent 
$\eta$-meson photoproduction reads
\begin{equation}
\label{fact}
{\sigma}_{\gamma \eta }^A=
{\sigma}_{\gamma \eta }^p
\frac   
{Z  + \zeta (A-Z)} { A} \times A_{eff}
\end{equation}
where
\begin{equation}
\label{aef}
A_{eff} = \frac {1} {2 \pi}
\int_0^{+\infty} d{\bf b}  \int_{-\infty}^{+\infty} dz \ {\rho}({\bf b},z) 
\int_0^{2\pi} d\phi \ exp \left[-{\sigma}_{\eta N}
\oint d\xi \ {\rho}({\bf r}_{\xi}) \right]
\end{equation}
Here ${\rho}(r)$ is the single particle density function, which
was taken of  Fermi type with parameters for each nucleus 
from~\cite{Jager}.
The last integration in Eq.~(\ref{aef}) being  over the path 
of the produced  $\eta$-meson 
\begin{equation}
r_{\xi}^2 = (b+ \xi cos\phi sin\theta)^2+(\xi sin\phi sin\theta )^2
+(z + \xi cos\theta )^2
\end{equation} 
Here $\theta$ is the emission angle of the $\eta$-meson 
relative to $\gamma$-momentum.

Eq.~(\ref{aef}) is similar to those from~\cite{Vercellin,Hufner} 
and in the low energy limit, i.e. by integration over the $\eta$-emission
angle $\theta$ becomes as~\cite{Benhar}
\begin{equation}
\label{ave}
A_{eff} =
\int_0^{+\infty} d{\bf b}  \int_{-\infty}^{+\infty} dz \ {\rho}({\bf b},z) 
exp \left( -{\sigma}_{\eta N}
\int_z^{\infty} d\xi \ {\rho}({\bf b}, \xi) \right)
\end{equation}
In the high energy limit, i.e. with the small angle scattering 
approximation $\theta =0$,  Eq.~(\ref{aef}) reduces to simple 
formula from~\cite{Margolis2}
\begin{equation}
\label{eq2}
A_{eff}=
\frac {1} {{\sigma}_{\eta N}}
\int_0^{\infty} d{\bf b} 
\ \left( 1- exp \left[ -{\sigma}_{\eta N} \ 
\int_{-\infty}^{+\infty} dz \ {\rho}({\bf b},z)
 \right] \right) 
\end{equation}

\begin{figure}[h]
{\psfig{figure=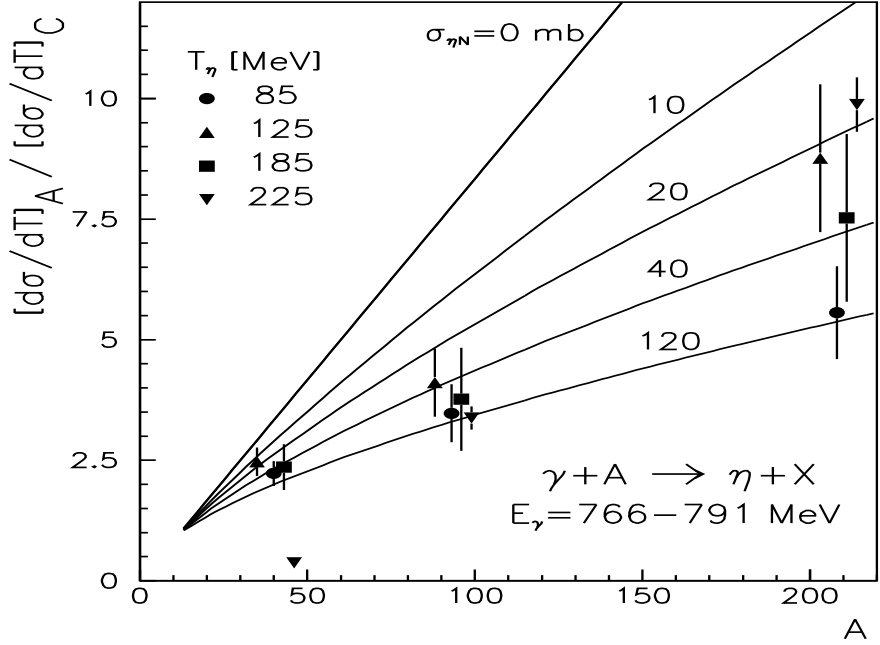,width=14cm,height=10cm}}
\caption{\label{fi1}The ratios of the differential cross sections
as function of the target mass and for different $\eta$-kinetic
energies. The experimental data are taken from~\protect\cite{Krusche}.
Lines show our calculations for several values of ${\sigma}_{\eta N}$.}
\end{figure}

The nuclear transparency is defined now as
\begin{equation}
T^A = \frac {{\sigma}_{\gamma  \eta }^A }
{{\sigma}_{\gamma \eta }^p \times \left[ Z  + \zeta (A-Z) \right]}
\end{equation}
and in the Glauber model it is simply given by
\begin{equation}
T^A = A_{eff} / A
\end{equation}
being the function of the target mass $A$, emission angle $\theta$
and in-medium $\eta N$ cross section ${\sigma}_{\eta N}$.
Note that this model neglect the in-medium effects~~{\it (i-iii)},
and takes only the final state interactions into account.

We analyze now the recent MAMI data~\cite{Krusche} on 
$\eta$-photoproduction
from $^{12}C$, $^{40}Ca$, $^{93}Nb$ and $^{207}Pb$ 
at $E_{\gamma}<$800 MeV in order to resolve 
the dependence~(\ref{aef}) with respect to the  target mass.
The $\eta$-production threshold on a free nucleon
lies at $\simeq$706~MeV, thus our analysis is expected to be valid for
$E_{\gamma} \geq 750$~MeV, in order to minimize effects~{\it (i-ii)}.
Moreover, to minimize the uncertainties related to~{\it (iii)},
which also are valid at high  $E_{\gamma}$, we  analyze the 
ratios of the differential cross sections integrated over the
$\eta$-meson emission angle as
\begin{equation}
\label{rat}
R(A/^{12}C) = \frac{d{\sigma}_{\gamma \eta }^A} {dT} 
\left( \frac {d{\sigma}_{\gamma \eta }^{^C}}{dT} \right)^{-1}
\end{equation}
We thus assume that the medium modifications of the
$N^{\star}$-resonance are almost the same for all nuclear targets.

The ratios~(\ref{rat}) are shown in Fig.~\ref{fi1} for several
kinetic energies of $\eta$-mesons and as a function of 
the target mass. The lines indicate our calculations
performed for different $\eta N$ cross sections.  The
model results are integrated over the $\theta$. Note that for 
${\sigma}_{\eta N}$=0 the ratio~(\ref{rat}) saturates at 
$R=A/C$ as was expected neglecting the final state interaction. 

\begin{figure}[h]
{\psfig{figure=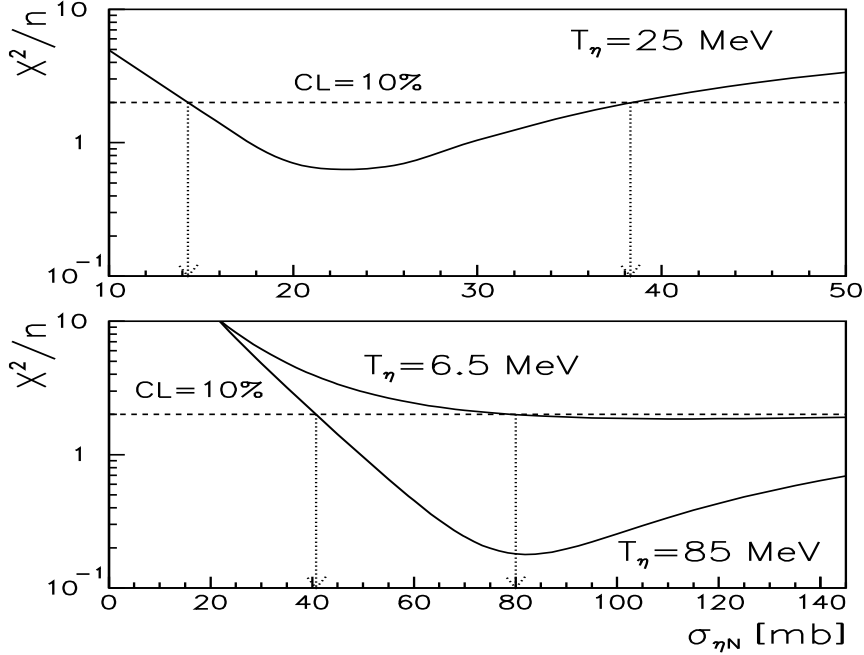,width=14cm,height=10cm}}
\caption{\label{fi2}The distribution of reduced ${\chi}^2$
as function of the $\eta N$ cross section.}
\end{figure}

We now fit the experimental ratios for each $T_{\eta}$
by minimizing the ${\chi}^2$ in order to evaluate
${\sigma}_{\eta N}$. A similar analysis was perfomed recently 
by Kharzeev et al.~\cite{Kharzeev} for the evaluation of the 
$J/\Psi$-nucleon
cross section. Fig.\ref{fi2} illustrates the
minimization procedure and shows sensitivity of the
data to the variation of $\eta N$ cross section.
We fixed the confidence level that gives  the value of
the reduced ${\chi}^2/n >2$ can be expected no more than 10\% 
of the time. With respect to the statistical errors of 
the experimental data
the minimization produces three types of results.
Namely, 1) with extraction of  ${\sigma}_{\eta N}$ and 
indication its uncertainty, 2) with
evaluation only the lower limit for ${\sigma}_{\eta N}$ or
3) with obtaining the minima behind the confidence level.  

\begin{figure}[h]
{\psfig{figure=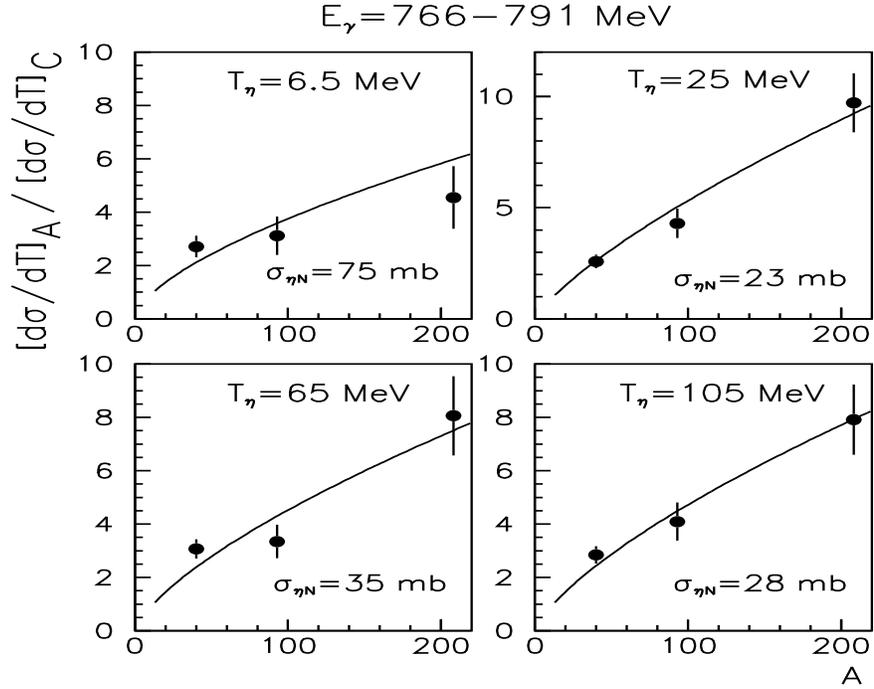,width=14cm,height=10cm}}
\caption{\label{fi3}The comparison between experimental 
data~\protect\cite{Krusche} and our fit.}
\end{figure}

\begin{figure}[h]
{\psfig{figure=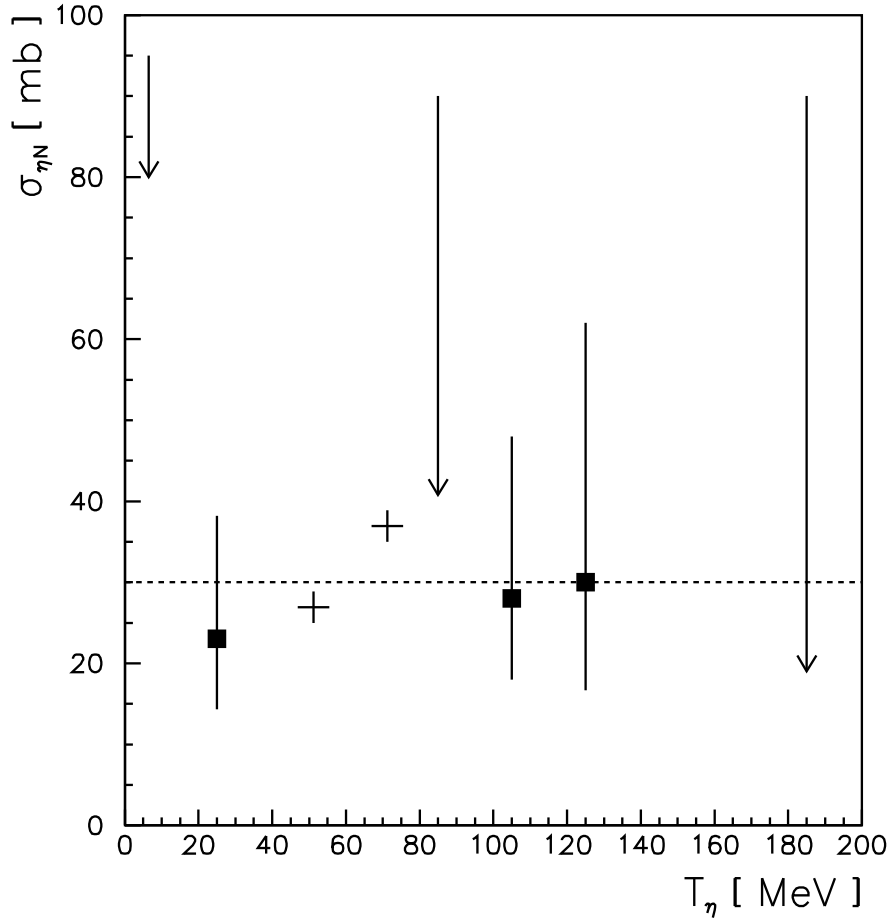,width=14cm,height=14cm}}
\caption{\label{fi4}Energy dependence of $\eta N$ cross section.
Squares indicate the results for the cases when we were able to
extract $\sigma_{\eta N}$ within 10\% confidence level, arrows when we could
only obtain a lower limit for $\sigma_{\eta N}$ and crosses if the result of 
our analysis is behind the confidence level.
The dashed line 
indicates a constant cross section of 30~mb.}
\end{figure}

Fig.~\ref{fi3} shows our final results in comparison with the 
experimental data and illustrate excellent agreement for 
wide range of the $\eta$-energies. Nevertheless we keep in
mind the uncertainties in evaluating of  $\eta N$ cross section
and collect the ${\sigma}_{\eta N}$ in Fig.~\ref{fi4} as
function of the $\eta$ energy and indication of confidence level.
Note that within present analysis we evaluate the inelastic 
(or absorption) $\eta$-nucleon cross
section, because the elastic scattering does not remove the
$\eta$-meson from the total flux, which was detected experimentally.

Our results indicate almost constant in-medium $\eta N$ cross section
as function of the $\eta$-energy in strong contradiction with 
the ${\sigma}_{\eta N}$ from the scattering in vacuum.   

To make a more definite conclusion about the suppression of the
$\eta N$ cross section in nuclear matter we need an accurate data on
the coherent $\eta$-photoproduction off nuclei. The coherent reactions
are more sensitive to the nuclear 
transparency ($\propto A_{eff}^2$~\cite{Bauer,Margolis2,Bochman})
and might solve the uncertanties of the present analysis
performed with the Glauber model.

\section{Results from BUU calculations}
\label{buu}
In order to verify the results from the previous section we use a BUU
transport model~\cite{Cassing1,Bertsch1,Teis1} to calculate energy 
differential $\eta$-photoproduction cross sections in nuclei. This
allows to drop several assumptions needed for the Glauber
calculations. Fermi motion and Pauli blocking are taken into account
as well for the primary $\eta$-production process as for the final state
interaction of the produced particles.

\renewcommand{\textfraction}{0.}
\renewcommand{\topfraction}{1.}  
\begin{figure}[t]
\centerline{
{\psfig{figure=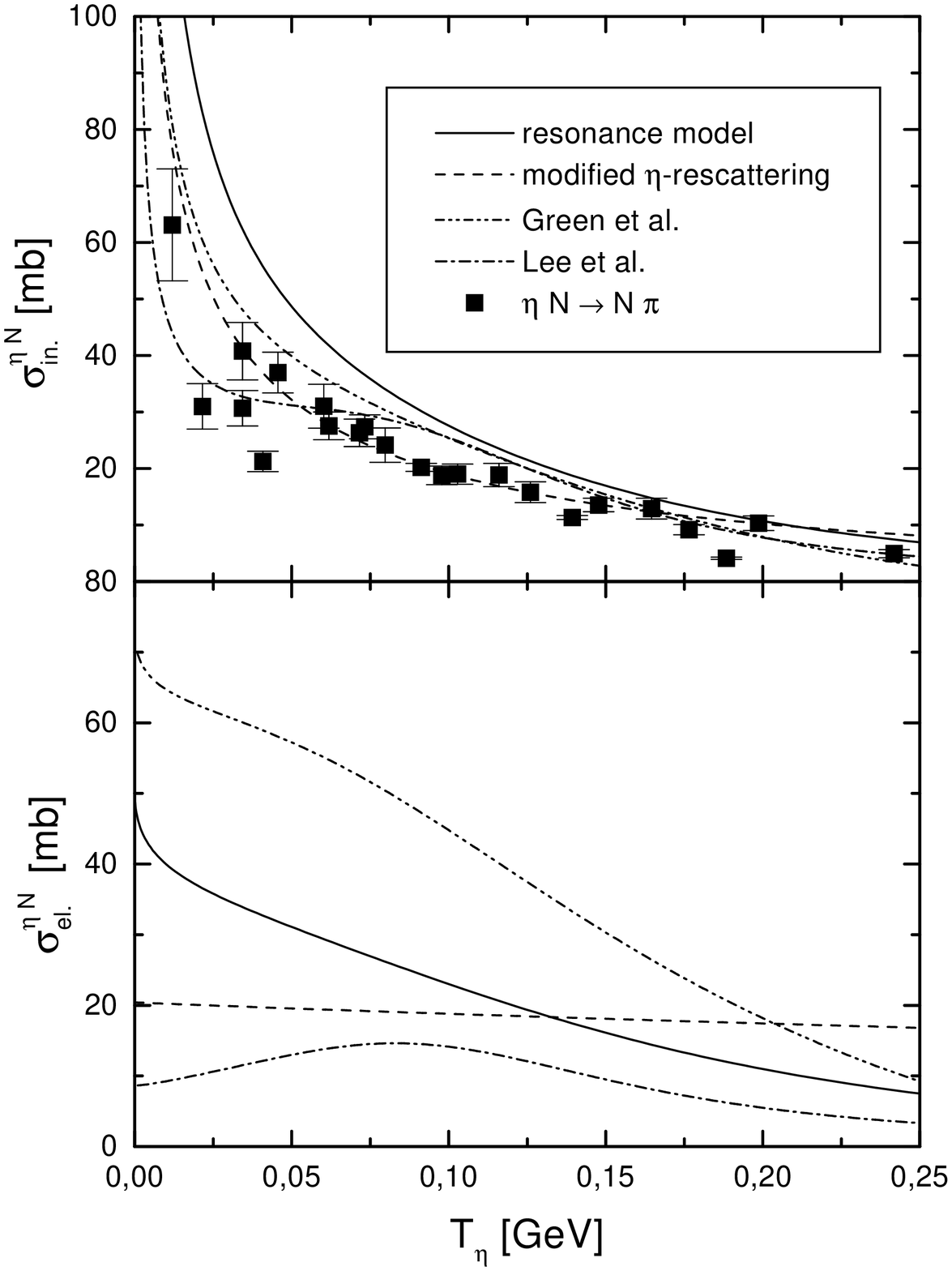,width=10cm}}}
\caption{\label{abs}Inelastic and elastic cross sections for 
$\eta N$ scattering with different models. Experimental data points are
obtained by detailed balance from $\pi^- p \to \eta n$ \protect\cite{Landolt}.
Note that the theoretical curves for the inelastic cross section can not 
directly be compared to the data because they contain additional channels.}
\end{figure}

\begin{figure}[t]
\centerline{
{\psfig{figure=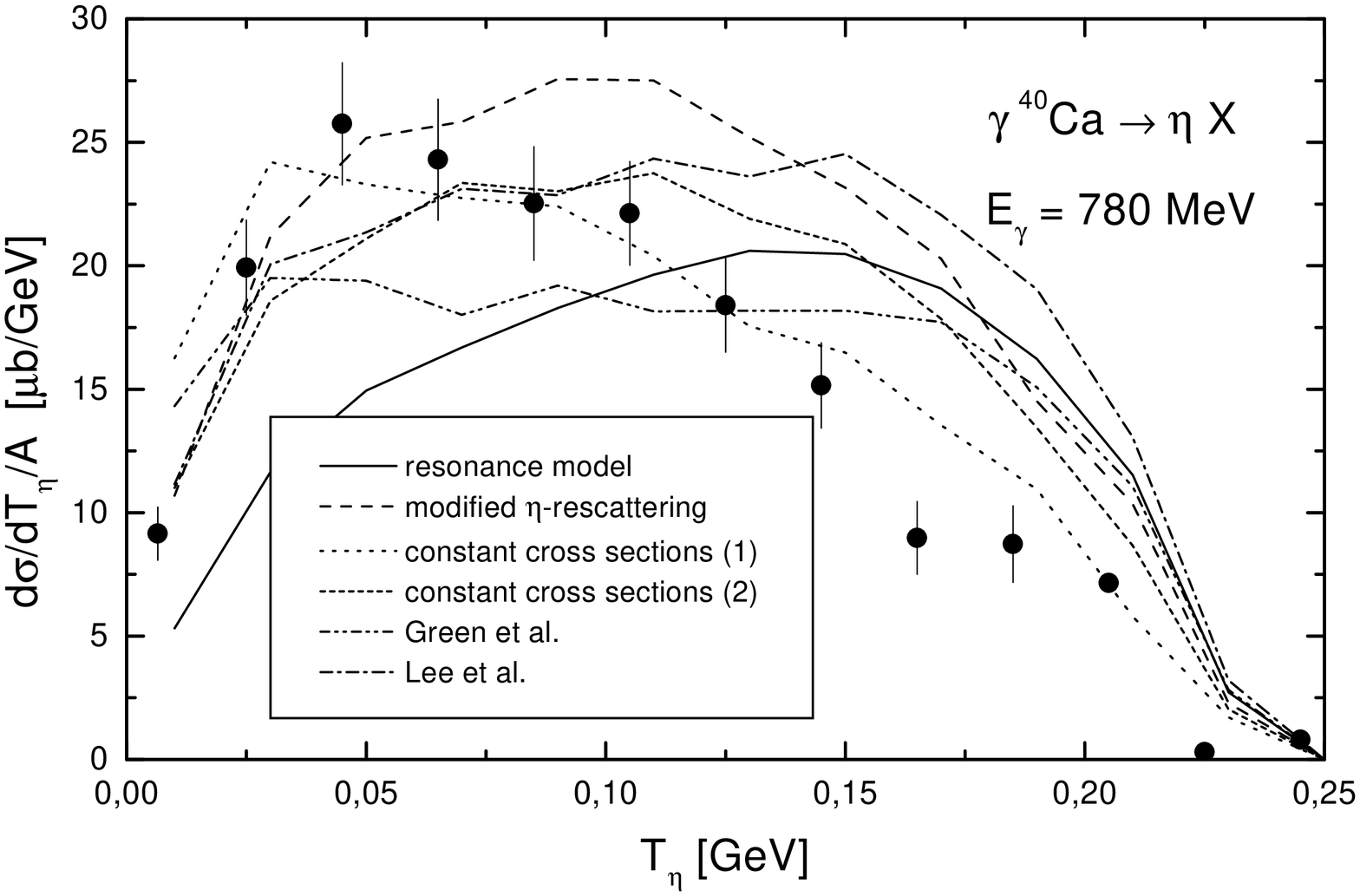,width=15cm}}}
\caption{\label{new}Energy differential $\eta$-photoproduction cross
sections for $E_\gamma=780\,$MeV in $^{40}$Ca.
Experimental data are from~\protect\cite{Krusche}. Lines indicate BUU 
calculations with $\eta N$ cross sections as
described in the figure. For further explanations see text.} 
\end{figure}

In Ref.~\cite{Effenberger} the BUU model was used to calculate 
$\eta$-photoproduction in nuclei with the resonance model for the
$\eta$ final state interaction from Ref.~\cite{Teis1}. Here the
$\eta$-rescattering was described by intermediate excitations 
of N(1535) resonances.   
The elastic and inelastic $\eta N$ 
cross sections calculated with this model
are shown in Fig.~\ref{abs} with the solid lines. It turned out that
this model was able to describe the total $\eta$-photoproduction cross
sections reasonably well but failed in the description of angular and
energy differential cross sections. Compared to the experimental data the
calculated cross sections were shifted to smaller angles and larger 
$\eta$-energies for all considered target nuclei and all 
photon energies up
to 800 MeV. In Fig.~\ref{new} the solid line shows the calculation of an
energy differential $\eta$-photoproduction cross section on 
$^{40}$Ca with this model.

It was already reported in Ref.~\cite{Effenberger} that the 
discrepancy to the experimental data is cured by using 
an energy independent $\eta N$ cross
section. The corresponding result is shown in Fig.~\ref{new} by the line
labelled 'constant cross sections (1)' where an inelastic cross section 
$\sigma^\eta_{in}=30\,$mb and an elastic cross section 
$\sigma^\eta_{el}=20\,$mb was used. 

Now we want to study the influence of different 
prescriptions for $\eta N$ scattering.    
The dashed line in Fig.~\ref{new}
indicates the results calculated with a modified 
$\eta$-rescattering model:
\begin{eqnarray}
\label{modi}
{\sigma}_{\eta N \to \pi N} &=& \frac{q_{\pi}}{q_{\eta}s} \
\frac{c^2 }{c^2+q_{\eta}^2} \ 40 \, {\rm mb \,GeV^2}
\, ,\, c=0.3\,{\rm GeV}  \\
{\sigma}_{\eta N \to \eta N} &=& \frac{45 \, {\rm mb\,GeV^2}}{s}    
\nonumber
\end{eqnarray}
where $q_{\pi}$, $q_{\eta}$ are the cms momenta of the
$\pi$- and $\eta$-meson, respectively, and $s$ stands for the squared 
invariant energy.  
The inelastic cross section was obtained 
by fitting the experimental
data for the reaction $\pi N \to \eta N$, while 
the elastic cross section was assumed to be equal 
to the one in the resonance
model at an $\eta$-energy of 125 MeV. The resulting elastic and inelastic 
cross sections
are shown in Fig.~\ref{abs} with the dashed line.
As can be seen from Fig.~\ref{new} (dashed line) this model improves 
the description of
the energy differential cross section for small $\eta$-energies but still
overestimates the cross section for higher energies. 

We have also used the $\eta N$ cross sections from Green and 
Wycech~\cite{Green} and Lee et al.~\cite{Lee}. 
The calculation of Green and Wycech is based on the K-matrix method and 
includes the $S_{11}(1535)$ and $S_{11}(1650)$ resonances. Following the
authors this model is valid up to an invariant energy of about 100 MeV
from $\eta N$ threshold which corresponds to an $\eta$ kinetic energy in
the nucleon rest frame of 160 MeV. Lee et al. use a parameterization of
the $\eta N$ scattering amplitude that is based on the calculation of
Bennhold and Tanabe~\cite{Ben}. This model contains the $P_{11}(1440)$,
$D_{13}(1520)$ and $S_{11}(1535)$ resonances and might therefore be limited
to an $\eta$ kinetic energy of about 100 MeV.

The total cross sections within these models
are shown in Fig.~\ref{abs}. Both models give about the same inelastic cross
section which is basically due to the fact that in both models the dominating
inelastic channel is given by the process $\eta N \to N \pi$. The 
experimental data~\cite{Landolt} for this reaction, obtained by detailed
balance from $\pi^- p \to \eta n$, are also shown. But one should note that
the theoretical curves contain additional, even though small, contributions 
from $\eta N \to N \pi \pi$ and
therefore can not directly be compared to these data points. The elastic 
$\eta N$ cross section in both models is very different which is an
indication for the large theoretical uncertainties in the models for
$\eta N$ scattering even in the vacuum.
 
The corresponding results of the BUU calculations for
photoproduction are given in Fig.~\ref{new}. 
As in the calculations within the resonance
model and the model from Eq.~(\ref{modi}) we again fail to describe the
shape of the energy differential cross section. The same holds for all
target nuclei and photon energies as well as for the angular differential
cross sections. Apart from the resonance model~\cite{Teis1} all models
give a satisfactory description of the cross section for $\eta$-energies
below 50 MeV. The failure of the resonance model is due to the fact that
this model was fitted to a larger class of elementary processes and a 
wider kinematical range and overestimated
the cross section for $\eta N \to N \pi$ in the considered energy range.   

In Fig.~\ref{new} we also show the result of a model calculation with a
constant inelastic cross section $\sigma_{in}^\eta=30\,$mb where we
neglected elastic $\eta N$ scattering (curve labelled 
'constant cross sections (2)'). Compared to the previous calculation with 
constant cross sections that included an elastic cross section the energy
differential cross section is shifted to larger energies and fails to
describe the data. Moreover the integrated cross section is slightly larger
because the elastic cross section increases, in average, 
the length of the path of
the produced etas through the nucleus and therefore reduces the number of
etas that escape from the nucleus. One sees that in our model a constant
inelastic $\eta N$ cross section alone is not sufficient to describe the
data but an elastic cross section is also needed.        
 
Since the models for $\eta N$ scattering in the
vacuum show a rather strong decrease of the total $\eta N$ cross section
with $\eta$-energy we are not able to reproduce the data for 
$\eta$-photoproduction with any of these models within our transport model
approach. For $\eta$-energies larger than
100 MeV we need an $\eta N$ cross section that is significantly larger than
the one from the vacuum models while for lower energies our calculations are
not very sensitive to the size of the cross section. A possible explanation
is that the vacuum models \cite{Green,Ben} are simply not applicable to
the considered energy range. After all, due to the Fermi motion of the
nucleons, the $\eta N$ cross section up to an invariant energy of 
$\sqrt{s}=1.74\,$GeV ($T_\eta=446\,$MeV in the nucleon rest frame) enters
the calculations for an eta with a kinetic energy of 250 MeV in the rest frame
of the nucleus.  
    
Our findings are in line with the Glauber analysis from 
section~\ref{glauber} and Ref.~\cite{Krusche} which need a constant inelastic
cross section of 30 mb in order to describe the mass dependence of the
energy differential cross sections.   
However, Lee et al.~\cite{Lee} 
were able to reproduce
energy differential cross sections  within the 
DWIA framework by using vacuum
$\eta N$ cross sections. One crucial 
difference to our calculation is that in
their calculation the outgoing nucleon in the elementary photoproduction 
process $\gamma N \to N \eta$ is set on-shell while in our semi-classical
treatment the elementary process takes place instantaneously with following
propagation of the produced particles through the nucleus. The potential
energy which is needed to set the nucleon on-shell clearly shifts the 
$\eta$-spectrum to lower energies. A priori it is not obvious which of the
two prescriptions is better suited to model the physical reality. Only a
DWIA calculation along the line of Ref.~\cite{Li} without the local
approximation of Ref.~\cite{Lee}
could clarify this question. An indication for a larger
$\eta N$ cross section at higher energies is the fact that we are able to
describe energy and angular differential cross section for 
$\eta$-photoproduction simultaneously by using a constant cross 
section~\cite{Effenberger} while in the calculations of Lee et al. the
angular differential cross sections are shifted to smaller angles compared
to the data.     
   
\section{Summary}
We have analyzed the $\eta$-photoproduction in nuclei
within the framework of the Glauber model and 
a BUU transport model approach. 

Using the standard Glauber theory we investigate
the $A$-dependence of the reaction $\gamma A \to \eta X$
in order to extract the data on $\eta$-meson final state
interaction in nuclei. It was found that the in-medium
$\eta N$ cross section is almost energy independent
from $\eta N \to \pi N$ threshold up to $\eta$-kinetic energy
of 130 MeV. 

Within a BUU transport model calculation we are able to reproduce
energy and angular differential data for $\eta$ photoproduction only by
using an energy independent $\eta N$ cross section but not with any
available $\eta N$ vacuum cross section. However, the effect 
of the nucleon
potential that can not be treated in our semi-classical calculation in a
correct way might have an impact on that conclusion.

\section{Acknowledgement}

The authors are grateful to B.~Krusche and H.~Stroeher
for productive discussions. They especially like to thank
U.~Mosel for valuable suggestions and a careful reading of
the manuscript.

\end{document}